\def \vc #1{{\mbox{\boldmath $#1$}}}

\def\thetag{{\vc \theta}}

\def\alphag{{\vc \alpha}}
\def\deltag{{\vc \delta}}
\def\gammag{{\vc \gamma}}

\documentclass[]{spie}  
\usepackage[dvips]{graphicx}

\title{Prospects for weak lensing/cosmic shear with  VLTs} 

\author{Mellier Y.\supit{a,b,c}, van Waerbeke L.\supit{a},
  Bertin E.\supit{a,b,c}, Tereno I.\supit{a,d}, \\
  P. Schneider\supit{e},  
 F. Bernardeau\supit{f} and 
T Erben\supit{e}
\skiplinehalf
\supit{a}IAP, 98 bis Blvd Arago, 75014 Paris, France \\
\supit{b}Obs. de Paris/LERMA, 77 Av. Denfert-Rochereau, 75014 Paris, France \\
\supit{c}TERAPIX-IAP, 98 bis Blvd Arago, 75014 Paris, France \\
\supit{d} University of Lisboa, Dept. of Physics, 1749-016, Lisboa, Portugal, \\
\supit{e} IAEF, Auf dem H\"ugel 71, 53121 Bonn, Germany \\
\supit{f}SPhT, CE Saclay, 91191 Gif sur Yvette cedex, France.
}

\authorinfo{Further author information: (Send correspondence to Y. Mellier)\\Y. Mellier: E-mail: mellier@iap.fr, Telephone: (+33) 1 44 32 81 40\\  
}

\begin{document} 
  \maketitle 

\begin{abstract}
The present status of weak lensing analyses of clusters of galaxies 
  and of cosmic shear surveys are presented and discussed. We focus on the 
  impact of very large telescopes on present-day and  future surveys
   and compare their potential with HST or  wide field 4 meter telescopes.

\end{abstract}



\section{INTRODUCTION}
Over the past decade, gravitational weak lensing offered astronomers 
  an observing tool capable to measure the amount and the spatial 
  distribution of dark matter in the sky.  Weak lensing analysis can 
probe distortion fields of any 
  specific gravitational systems, like groups or clusters of galaxies, in
  order to provide individual mass reconstruction of those systems. 
  Cosmic shear studies, on the other hand, probe the statistical properties of 
   the gravitational shear, relate them to the geometry and matter/energy
  content of the universe and attempt to measure some cosmological parameters 
   and the dark matter power spectrum.  The results obtained recently are 
  reviewed in (see \citenum{fortmel94}, \citenum{mel99},
  \citenum{bs01},  or \citenum{melspie1} in this
  conference) and some of them will be presented in this proceeding. \\
In this presentation, we  only focus on the impact the VLT can have 
  on weak lensing science.  In order to answer this question, we first 
    sumarise the present status of  weak lensing and cosmic shear
  surveys, then  
  we point out the main issues and future goals and finally we try 
  to answer how can VLT help in the future.

\section{Gravitational weak lensing}
\label{sect:intro}  
The distortion of light beams produced by gravitational lensing 
   modifies the image properties of lensed galaxies.
  In a Friedman-Robertson-Walker metric,
   and for stationary and weak gravitational
fields, the deflection angle  writes
\begin{equation}
  \hat\alphag
  = \frac{2}{c^2}\,\int\nabla_\perp\Phi\, \ {\rm d}l \ ,
\label{deflec}
\end{equation}
where $c$ is the celerity and $\Phi$ the
  3-dimension gravitational potential associated to the lens.
   When the deflection angle is small and when lenses can be approximated as thin
   gravitational systems, the relation between
  the source ($S$) and its image ($I$) positions follows the simple
  geometrical ``lens equation'':
\begin{equation}
\thetag^I=\thetag^S+{D_{LS} \over D_{OS}} \hat\alphag(\thetag^I) \ ,
\label{lensequa}
\end{equation}
where $D_{ij}$ are angular diameter distances.\\
Equations (\ref{deflec}) and (\ref{lensequa}) express how
  lens properties depend on (dark) matter distribution and
  on cosmological models.
  From an observational point of view, gravitational lenses
  manifest as image multiplication of galaxies or quasars, strong and weak
    distortions of galaxy shape or
  transient micro-lensing effects.  These effects, as well as 
   the  time delays attached to image multiplications 
 are exploited
 in order to  probe the geometry and matter/energy content 
  of the Universe or to observe high-redshift 
galaxies
 (see \citenum{fortmel94}, \citenum{mel99}, 
  \citenum{bs01}, \citenum{blandnar92} for reviews).
\\
The image magnification is  characterized by the 
   convergence, $\kappa$,  
   and by the shear components $\gamma_1$ and $\gamma_2$:
\begin{equation}
 \kappa = \frac{1}{2}\,(\varphi_{,11}+\varphi_{,22})  ; \ \  \ \ \ \ \
  \gamma_1(\theta) =
  \frac{1}{2}\left(\varphi_{,11}-\varphi_{,22}\right) ; \ \ \ \ \ \ \
  \gamma_2(\theta) = \varphi_{,12} = \varphi_{,21}
\label{convshear}
\end{equation}
 {\rm where } the $\varphi_{,ij}$ are the second derivatives 
  of $\varphi$ with respect to the $i,j$ coordinates and
\begin{equation}
\varphi(\theta) = \frac{2}{c^2}\ \frac{D_{LS}}{D_{OS}D_{OL}}\,
  \,\int\,\Phi(D_{OL}\theta,z)\, \ {\rm d}z\;.
\label{projpoten}
\end{equation}
Note that $\kappa$ is the Laplacian of $\varphi$, so it is 
 the projected mass density.
\\
The shear applied to lensed galaxies
  increases  their ellipticity along a direction perpendicular to
  the gradient of the projected potential.  The lens-induced distortion
   $\deltag$ can then be 
    evaluated from the  shape of galaxies as it can be observed from
  the components of their  surface brightness second moment 
   $M_{ij}= \displaystyle{
 {\int
I\left(\thetag\right) \theta_i \ \theta_j
 \ {\rm d}^2\theta \over \int I(\thetag) \ {\rm d}^2\theta }} \ :
$
\begin{equation}
\deltag ={2 \gammag \ (1-\kappa)
\over
(1-\kappa)^2+\vert \gamma \vert^2} \  \  \ 
\Longrightarrow \vert \deltag \vert = {a^2 -b^2 \over a^2 + b^2 }  \ , 
\label{distor}
\end{equation}
where $a$ and $b$ are the major and minor axis of galaxies, as derived from 
  their surface brightness second moment. 
\\
In the weak lensing regime, 
  the relation between the distortion and the gravitational shear  
 simplifies to $\deltag \approx 2 \gammag $, so
 in principle, if we assume that unlensed galaxies are randomly 
oriented ({\sl ie} $\left< \varepsilon\right>^S=0$), the 
     ellipticity of galaxies, as measured
from the  second moment components, provides a
 direct estimate of the gravitational shear.
   The
  ``shear map'' can then be used to reconstruct the ``mass map''
 at any galaxy position.
     However,  since each galaxy has
its  own intrinsic ellipticity, and since the background galaxies
   only sparsely sample the sky, the final shear-induced
   ellipticity map is contaminated by 
  shot noise so galaxy ellipticity must be averaged over a 
  small but finite angular-size.
\begin{figure}[t]
\begin{center}
\includegraphics[width=11cm]{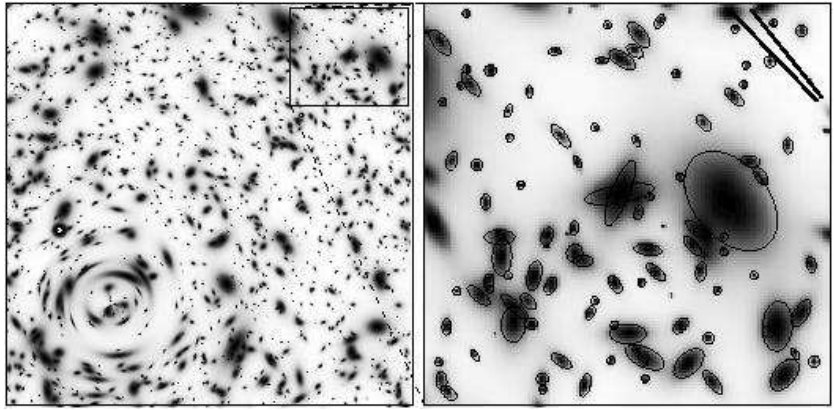}
\end{center}
\caption[]{The left panel shows a simulation of a cluster of galaxies at redshift 0.15,
 lensed by an isothermal sphere with a velocity dispersion of
1300 kms$^{-1}$.  The lensed population has an average redshift of one.
 The amplitude of the gravitational shear  decreases with radial
  distance and far from the cluster
center, the ellipticity produced by the shear is 
 smaller than the intrinsic ellipticity
 of the galaxies. The  lensing
 signal must be averaged over a large number of galaxies in order to
be measured
accurately. The zoom on the right panel shows the images of the galaxies in the
 weak lensing regime. The contours show their shape
as determined from their second moments. The average orientation of these
galaxies is given by
the solid lines at the top right. The lower line is the true orientation
of the shear produced by the cluster at that position and the upper line
is the orientation computed from the galaxies of the zoomed area. The
difference between the
two orientations is random noise due to the
   intrinsic ellipticity and orientation on
 distributions of the galaxies (from \citenum{mel99}).}
\label{sumlcluster}
\end{figure}
\section{Analysis of clusters of galaxies}
\label{sec:clusters}
\subsection{Weak lensing analysis of distant clusters}
\begin{figure}[t]
\begin{center}
\includegraphics[width=9cm]{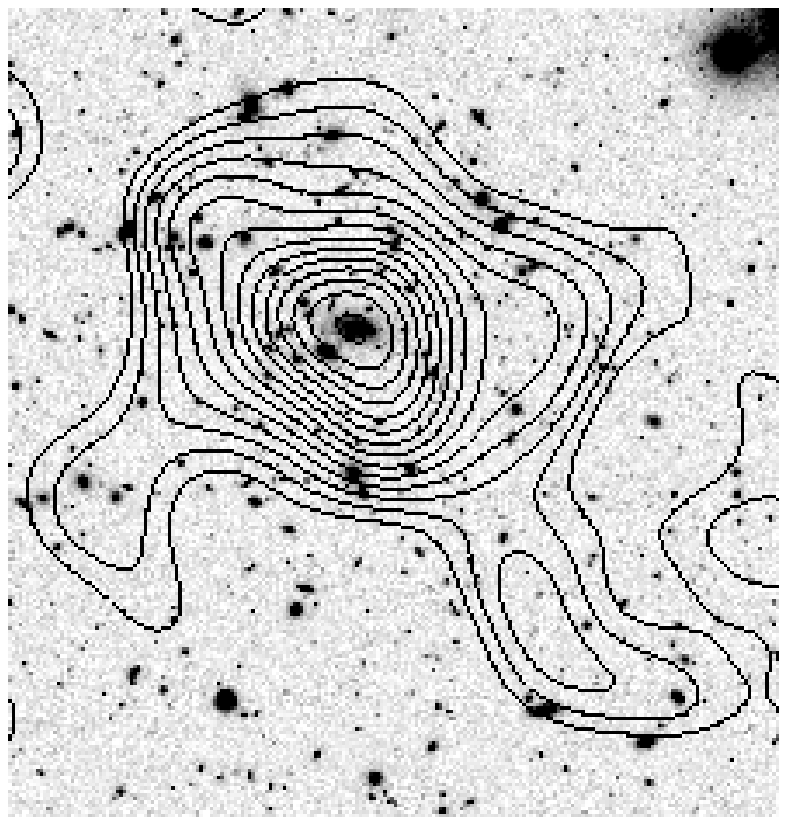}
\end{center}
\caption[]{A mass reconstruction obtained with VLT/FORS images. The 
clusters Cl1232-1250 ($z=0.5$) is part of the EDICS cluster sample 
  which contains 10 clusters at $z\approx 0.5$ and 10 at $z \approx 0.8$
 (White et al and Clowe et al in preparation; courtesy D. Clowe).
}
\label{clowemass}
\end{figure}

Because clusters of galaxies are young and the 
  largest relaxed  gravitational systems they 
  are surviving witnesses of the cosmic history
of structure formation.  Their present-day mass function, its
evolution with look-back time and their radial  mass density profile
  are remaining footprints of the growth 
  of gravitational perturbations over the past 
  Gigayears as well as physical illustrations of the roles of 
   cosmological parameters and the power spectrum of
 primordial fluctuations in the cosmic scenario.   \\
It is now well known that gravitational lensing 
  studies of clusters of galaxies do not need to assume 
   the geometry of the mass distribution nor its dynamical and thermodynamical
properties.  In this respect   
  this is an interesting tool to study complex systems like clusters
 of galaxies.
 By comnining the previous equations, it is easy to show that the 
  projected mass density can be reconstructed from this integral:
\begin{equation}
\kappa(\thetag)= {1 \over \pi}  \int \Re[\mathcal{D}^*(\thetag-\thetag')
\gammag(\thetag')]  \ d^2\theta' \ + \kappa_0  \ ,
\end{equation}
where
\begin{equation}
\mathcal{D}(\thetag-\thetag')={(\theta_2-\theta_2')^2 -
(\theta_1-\theta_1')^2- 2i
(\theta_1-\theta_1')(\theta_2-\theta_2') \over
\vert (\thetag-\thetag') \vert^4 } \ .
\end{equation}
This inversion produces the mass reconstruction from the shear field.
\\
From a technical and practical point of view
clusters of galaxies are  attractive  systems
  for weak lensing analysis.  Their mass-density contrast is high
($>100$) enough to produce significant
  gravitational distortion, and their angular scale
  ($\ge $ 10 arc-minutes) is much larger than the typical
  angular distances between lensed background galaxies.
 The details of cluster mass distribution can therefore be
 well sampled by the grid of lensed galaxy population.
\\
Over the past decade about 50 clusters have been reconstructed
using weak gravitational lensing analysis, which all produced 
  mass maps as the one shown in Fig.(\ref{clowemass}). They are listed
  in Table \ref{tabwl} (I excluded nearby clusters and reconstruction
  from depletion by magnification bias).
     This table reveal some  general trends.
  The averaged  mass-to-light ratios from
  weak lensing ($WL$) is $\left(M/L\right)_{WL}\approx 400 \ h$ and typical
velocity dispersion of those clusters is 1000 km/sec.
   When scaled with respect to the critical mass-to-light ratio,
  an estimate of the mass density of the universe can be derived,
\begin{equation}
\Omega_{m-WL} \approx 0.3 \ \pm 0.1 \ ,
\end{equation}
  which is in very good agreement with values inferred by other 
  methods.
\begin{figure}[t]
\begin{center}
\includegraphics[width=6cm,angle=270]{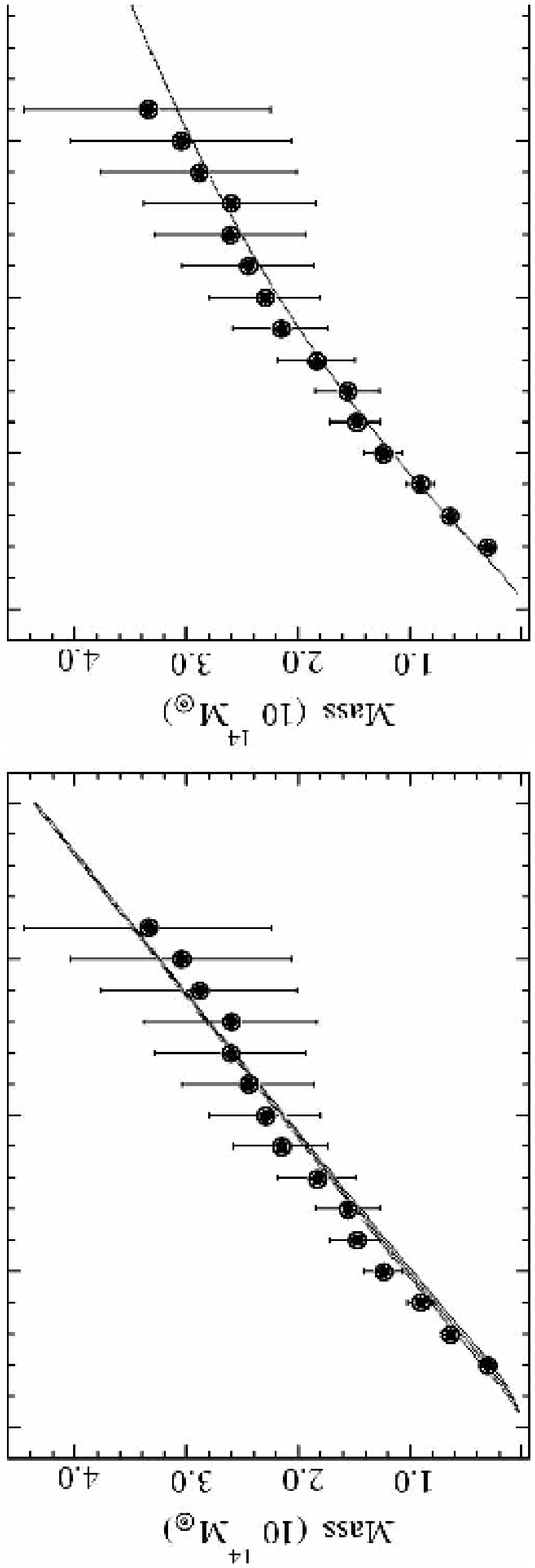}
\end{center}
\caption[]{The total mass profile of the lensing cluster MS1007-1224 as
 it is derived from a mass reconstruction of VLT FORS/ISAAC and NTT/SOFI 
  data.  The solid lines on the left show the best fit of an singular isothermal
  profile and an isothermal model with core radius.  The solid line on the 
  right in an NFW mass profile.  Both isothermal and NFW profiles fit 
  these data equally well (from \citenum{athreya02}, see also
  \citenum{lomb00}.).
}
\label{mas1008radialmass}
\end{figure}
In contrast, the mass density profiles are still too noisy to 
  provide robust  parameter profiles, like its central value, its typical 
  scale and its slope.  Isothermal, power law or NFW
  ``universal'' models fit equally well
    the data (see Fig.(\ref{mas1008radialmass})), for all clusters analyzed so
far (see \citenum{clo00}). It is therefore not yet possible to test
   current
  numerical CDM predictions using this method. In
fact, in view of the present signal-to-noise ratio of mass maps and
  the large family of possible analytical mass profiles, we are
 still far from being able to address in details this issue, if
  one use only weak distortion analysis.
\begin{table}
{\small
\caption{Results obtained from weak lensing analyses of
 clusters. The
scale is the typical radial distance with respect to the cluster center.
 The (M/L)$_r$ ratio has been rescaled since most data where
obtained with different filters (this is a rough rescaling; it probably
   increases uncertainties).
}
\label{tabwl}
\begin{tabular}{lccccccl}\hline
Cluster& $z$ & $\sigma_{obs}$ &  $\sigma_{wl}$ & M/L & Scale& Tel. &
Ref. \\
 & & (kms$^{-1}$)  & (kms$^{-1}$)& ($h_{100}$) & ($h^{-1}_{100}$ Mpc) &
&  \\
\hline
A2218 & 0.17 & 1370 & 1000-1400& $\approx 300$ & 0.5 & CFHT & Squires et al 
(\citenum{squi96a})\\
 &  &  &-& 310 & 0.1 & HST & Smail et al (\citenum{smail97})\\
A1689 & 0.18 &2400&  1200-1500 & - & 0.5 & CTIO & Tyson et al (\citenum{tvw90})\\
      &  & &-& 400& 1.0 & CTIO & Tyson \& Fischer (\citenum{tysfis95})\\
      &  & &1030& -& 1.0 & ESO/2.2 & Clowe \& Schneider (\citenum{closhn01})\\
A2163 & 0.20 & 1680&740-1000  & 300&0.5 &  CFHT & Squires et al
(\citenum{squi97})\\
A2390 & 0.23 & 1090& $\approx$1000 &320 & 0.5&CFHT & Squires et al
(\citenum{squi96b}) \\
 $<8  \ clusters>$ & $<0.2>$ & - & & $<295>$ & 1.0 & CTIO & Wittman et al (\citenum{wit00b})\\
MS1455+22 & 0.26  &1133 & - &1080&0.4 & WHT & Smail et al
(\citenum{smail95})\\
AC118 & 0.31 & 1950 & -&370 & 0.15 &HST  &  Smail et al (\citenum{smail97})\\
MS1008-1224 & 0.31 & 1054 & 940 & 340 & 0.5 & VLT & Athreya et al (\citenum{athreya02})\\
 &  &  & 850 & $\approx $ 320 & 0.5 & VLT & Lombardi et al (\citenum{lomb00})
\\
MS2137-23 & 0.31 & - &  950 & 300 & 0.5 & VLT & Gavazzi et al (in prep.) \\
MS1358+62 & 0.33  & 940&780 &180 & 0.75&HST  & Hoekstra et al (\citenum{hoek98})\\
MS1224+20 & 0.33  & 802& -& $\approx$ 800& 1.0& CFHT & Fahlman et al
(\citenum{falh94})\\
 &   & & 1300 &  890& 1.0& MDM2.2 & Fischer (\citenum{fisch99}) \\
Q0957+56  & 0.36  & 715&- & -& 0.5& CFHT& Fischer et al (\citenum{fisch97}) \\
Cl0024+17 & 0.39  & 1250 &-& 150& 0.15 & HST &  Smail et al (\citenum{smail97})\\
          & & & 1300& $\approx$900& 1.5& CFHT & Bonnet et al (\citenum{bon94})\\
Cl0939+47 & 0.41  & 1080 &-&120 & 0.2 & HST &  Smail et al (\citenum{smail97})\\
          &    &  &-&$\approx$250 & 0.2& HST& Seitz et al (\citenum{seitz96}) \\
Cl0302+17 & 0.42  & 1080& & 80& 0.2 & HST &  Smail et al (\citenum{smail97})\\
RXJ1347-11 & 0.45  & - &1500& 400& 1.0 & CTIO& Fischer \& Tyson
(\citenum{fischtys97})\\
3C295 & 0.46 & 1670 &1100-1500 &- &0.5 & CFHT & Tyson et al (\citenum{tvw90})\\
 &  & &-& 330&0.2 & HST &   Smail et al (\citenum{smail97})\\
Cl0412-65 & 0.51  &-&-&70 & 0.2& HST & Smail et al (\citenum{smail97}) \\
Cl1601+43 & 0.54  &1170 &-&190 & 0.2& HST & Smail et al (\citenum{smail97}) \\
MS0016+16 & 0.55  &1230 &-&180 & 0.2& HST &  Smail et al (\citenum{smail97})\\
          &       & &740 &740&0.6 & WHT & Smail (\citenum{smail93})\\
          &       & &800 &-&0.6 & Keck & Clowe et al (\citenum{clo00})\\
MS0451 & 0.55  &1371 &980&- & 0.6& Keck & Clowe et al (\citenum{clo00})\\
Cl0054-27 & 0.56  &-&-&400 & 0.2& HST &  Smail et al (\citenum{smail97})\\
MS2053 & 0.59  &820&730&- & 0.5& Keck &  Clowe et al (\citenum{clo00})\\
 &   &&886&360 & 0.5& HST &  Hoekstra et al (\citenum{hoekstra02b})\\
MS1137+60 & 0.78  &884&1190&270 &0.5& Keck&  Clowe et al (\citenum{clo98},\citenum{clo00})\\
RXJ1716+67 & 0.81  &1522&-&190 &0.5 &Keck & Clowe et al (\citenum{clo98})\\
 &   & & 1030&- &0.5 &Keck & Clowe et al (\citenum{clo00})\\
MS1054-03 & 0.83 & 1360 & 1100-2200 &350-1600 &0.5 & UH2.2& Luppino \&
 Kaiser (\citenum{lupkais97})\\
 &  & & 1310 &250-500 &0.5 & HST & Hoekstra et al (\citenum{hoek00}) \\
 &  & & 1080 &- &0.5 & Keck & Clowe et al (\citenum{clo00}) \\
$<10  \ clusters>$ & $<0.5>-$ & - & &  & 1.0 & VLT & White et al (2002)\\
 EDICS & $-<0.8>$ & - & &  &  &  & and Clowe et al (2002)\\
\hline
\end{tabular}
}
\end{table}
\begin{figure}[t]
\begin{center}
\includegraphics[width=6cm,angle=270]{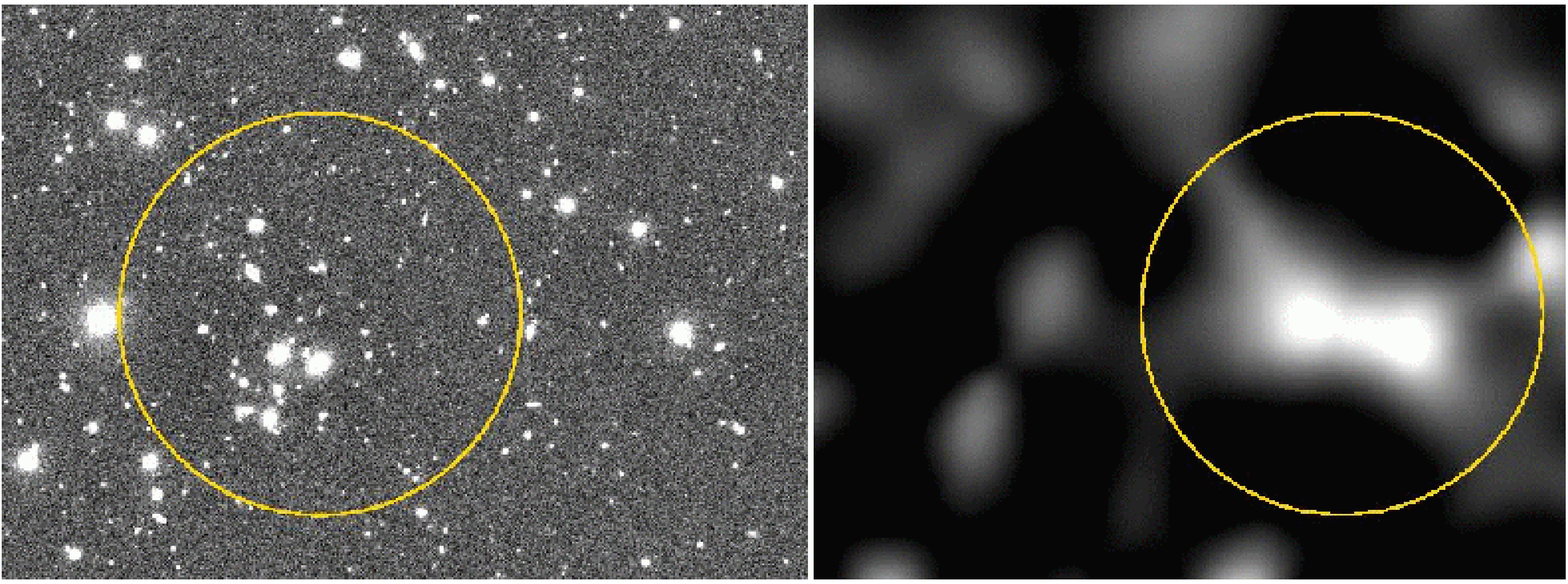}
\end{center}
\caption[]{A blind mass reconstruction in a randomly selected VLT field. 
  The right panel shows the $\kappa$ mass automatically produced 
   by the cosmic shear pipeline used by (\citenum{mao01}) to process
 their 50 VLT fields. After this strong detection, this field was 
  checked by eye and a galaxy concentration corresponding to a cluster 
  of galaxies was detected. This show the strength on weak lensing analysis: 
   in principle, this technique allows astronomers to make 
  mass-selected cluster samples without using the light distribution 
   of galaxies (from \citenum{maolimessenger}).
}
\label{maolicluster}
\end{figure}
\subsection{Dark clusters?}
One of the interesting outcomes of weak lensing is the 
  blind discovery of clusters, like in Fig.(\ref{maolicluster}) 
  (see also \citenum{wittman01}), or the remarkable discovery of
dark cluster candidates. Dark clusters are 
  massive objects which are
   only detected from gravitational lensing effects. The most
   convincing cases (HST/STIS field
 (\citenum{mirallesdark}); 
   Abell 1942 (\citenum{erben00}); Cl1604+4304,
  (\citenum{umet00}); Cl0024+1654, \citenum{bon94})
  have been detected from a strong shear field spread over
more than one arc-minute.  For almost all
   candidates,  there is no galaxies associated to the lens\footnote{The
  restriction is for MS1224 reported in Table \ref{tabwl}.}.
   Their typical mass,  estimated from reasonable assumptions
  on their redshift, corresponds to $M/L >> 500$.
  \\
There is not yet any  clear understanding about  ''dark clusters''.
  Abell 1942,  has also been observed
 in  the infrared in order to detect a very high-redshift cluster
  of galaxies (\citenum{gray01}) which could be missed on visible
  data.  But nothing has been detected so far.  It is
 premature to claim that these systems are totally
 dark until  X-ray observations confirm they do not contain
   hot intra-cluster gas.
 But if those objects were confirmed
  to be dark gravitational systems 
 that will be a serious theoretical issue in order to 
    understand how gravitational collapse would accrete
 only non-baryonic dark matter inside clusters. 
   If the  dark cluster population is numerous, then  
  X-ray or optically selected samples
  underestimate their mass fraction and the cluster abundance in the universe.
   Whether they contribute significantly
to $\Omega_m$ has to be clarified. However, 
  since only few dark clusters have been observed in wide field 
  surveys, it is likely that they do not represent a significant 
  fraction of mass.  This is also confirmed by the fact that 
   the $\Omega_m-\sigma_8$ relations inferred from cosmic shear (see
next section) and from  cluster abundance are not in  contradiction, as
  it should be if a significant fraction of clusters were dark.

\subsection{Issues and prospects for VLTs}
The investigation of a large sample of clusters of galaxies using weak 
  gravitational lensing is motivated by the potential of this method to
  directly probe matter, regardless its nature nor its dynamical stage. 
  In principle, any mass-scale can be observed, so that a detailed mass 
  function of gravitational systems could be derived from this kind of survey. 
However, there are still technical issues have to be addressed.
 Some  of these key questions are listed below :
\begin{itemize}
\item What is the lower mass limit weak lensing can probe?
\item Is it possible to build a complete mass-limited sample of clusters
of galaxies with a weak lensing survey?
\item Are dark clusters real? 
\item Could projection effects contaminate weak lensing reconstruction? How
  projection effect 
  could spoil the individual mass maps and the definition of a
   mass-limited sample?
\item What is the cluster mass profiles of clusters of galaxies and groups? 
  Down to what accuracy can we measure it with weak lensing?
\item What is the redshift distribution of sources? Do we have a good
  absolute mass calibrations of clusters?
\item Can we use weak lensing to probe substructures in cluster?
\item How can we suppress the mass-sheet degeneracy?
\item What is the mass of the most massive high-$z$ clusters of galaxies?
\end{itemize}

\begin{figure}[t]
\begin{center}
\includegraphics[width=7.2cm]{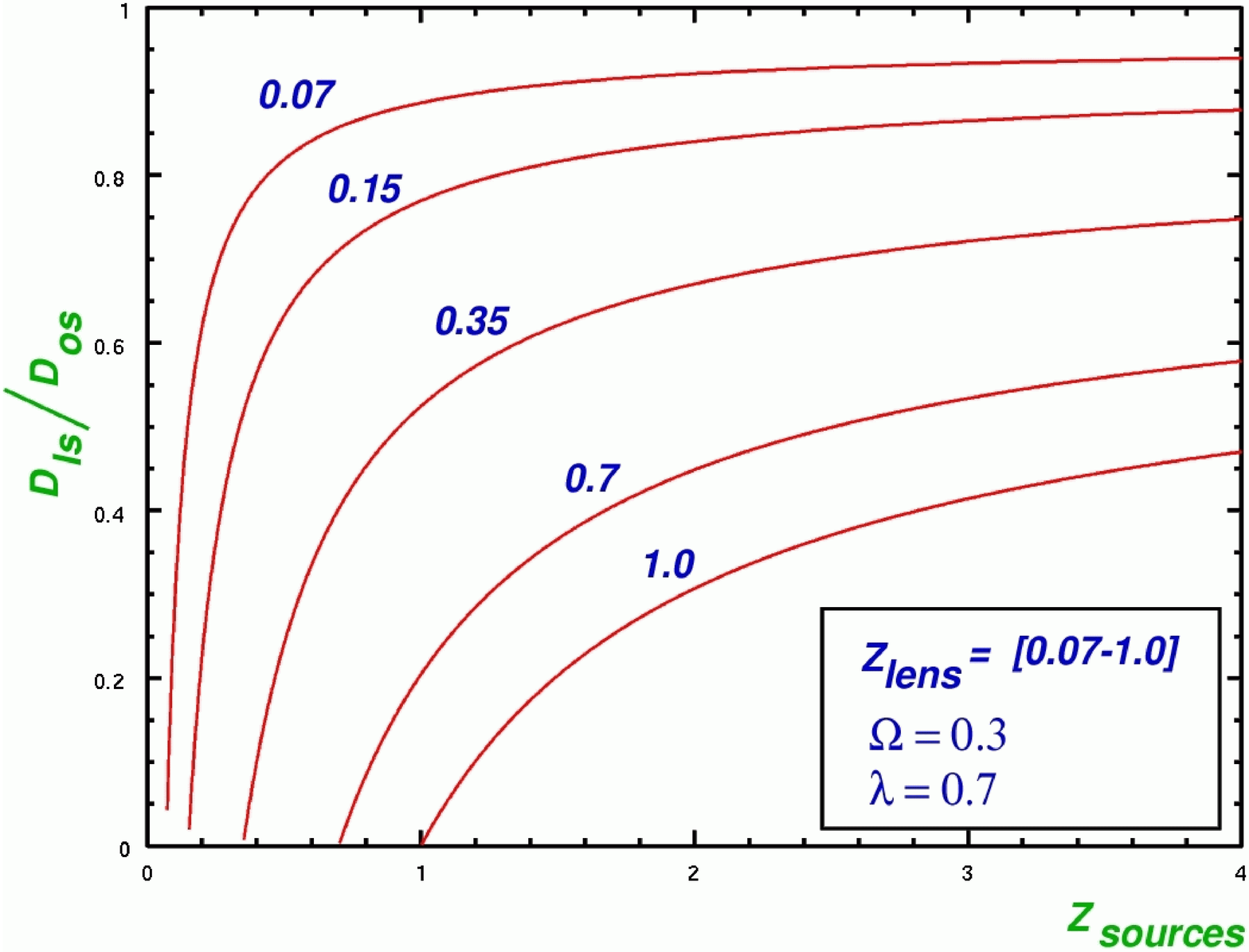}
\includegraphics[width=8cm]{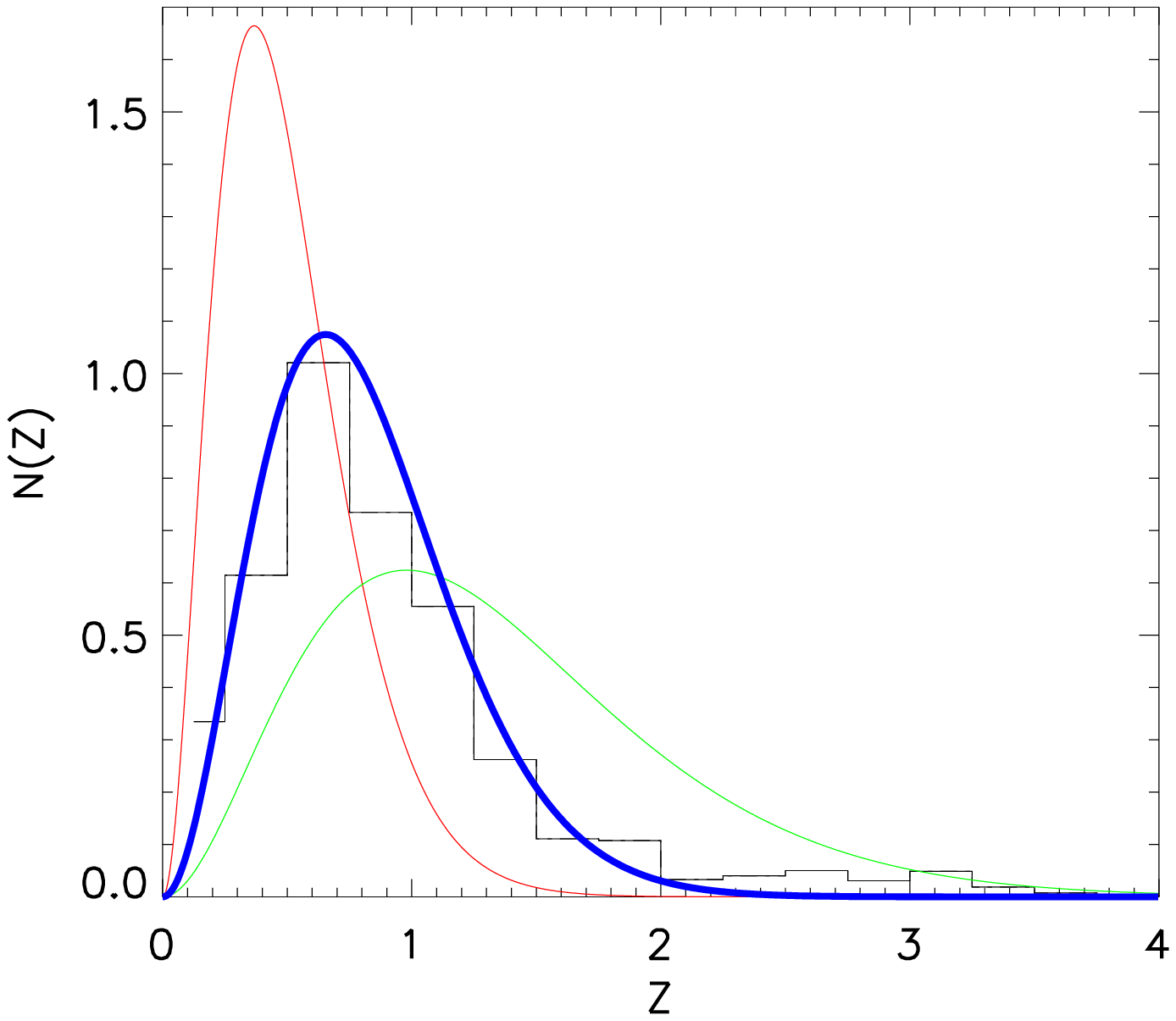}
\end{center}
\caption[]{Redshift distribution of sources.  The left panel shows 
  the sensitivity of the $D_{LS}/D_{OS}$ ratio to the 
  source and lens redshifts. For a fixed lens redshift, the 
 it monotonically increases with the source redshifts.
  Conversely, for a fixed source redshift, the ratio decreases 
  with the lens distance, making mass reconstruction of very 
distance cluster more and more challenging. On the right panel, the
  histogram shows the photometric redshift distribution of
  the van Waerbeke et al (\citenum{vw02}) cosmic shear shear survey. This photo$z$ 
  distribution was obtained using 1000 galaxies observed with  VLT
   FORS1, ISAAC and  NTT/SOFI. The thick line is the best fit.
}
\label{zsource}
\end{figure}
The VLT can certainly help to answer this questions.
 Although no ground based telescopes can compete with the outstanding image
  quality of HST, Keck, VLT, Gemini and SUBARU have much larger fields of 
  view which better match cluster angular sizes. 
\\
In order to emphasize the strength of VLT, the relevant quantity is the 
  signal-to-noise
 of the weak lensing analysis of a gravitational system.  For a typical 
mass reconstruction it expresses as follows:
\begin{equation}
{S \over N} \approx 10 \ \left[{n \over 20 \ arcmin^{-2}} \right]^{{1 \over 2}}
 \times \left[{ \sigma_{\epsilon_{gal}} \over 0.4} \right]^{-1}
  \times \left[{ \sigma_{{\rm v}} \over 800 \ km.sec^{-1}}\right]^{2}
 \times \langle {D_{LS} \over D_{OS}} \rangle   \ ,
\label{sncluser}
\end{equation}
 where $n$ is the galaxy number density, $\sigma_{\epsilon_{gal}}$ the
  intrinsic ellipticity dispersion of galaxies and $\sigma_{{\rm v}}$ the
 cluster velocity dispersion.  
Equation (\ref{sncluser}) shows that we only have two ways to improve 
  weak lensing analysis of a specific target: either increasing the depth of 
  observation in order
 to increase $n$, or increasing ${D_{LS} / D_{OS}}$ by  
 increasing the source redshifts and therefore the amplitude of the shear.  
  In addition,  since the ``Universal mass profile'' is only an 
   averaged view of cluster radial mass distribution, we expect 
  strong mass profile fluctuations from cluster to cluster, so it is  
  important to observe a large sample.
\\
 $n$ and  ${D_{LS} / D_{OS}}$ are
 correlated, since the redshift of sources increases with the 
   depth of observations. However,  both work in several and somewhat 
  different ways: $n$ also  
   provides a better sampling of each field, making the analysis of 
  substructures in clusters easier. On the other hand, 
  as it is shown on Fig.(\ref{zsource}), changing ${D_{LS} / D_{OS}}$ is more
  complicated since it both depends on the lensed sources and the distance
  of the lensing cluster. Nevertheless, for distant clusters of galaxies 
  the need for very high redshift sources is obvious. Furthermore, 
  it is worth noticing that the absolute mass of clusters cannot be 
  derived if the redshift of the lensed sources is unknown.
\\
The impact of VLTs are therefore clear: in order to improve the 
  spatial resolution of mass reconstruction, to produce mass maps 
  of very distant clusters of galaxies and to uncover their absolute 
  total mass, one need to go {\sl very } deep and to get redshifts 
  of galaxies. The need for redshift  is 
  a  critical point and indeed a very strong case for VLT 
  observations rather than 4-meter class telescopes: beyond $I \approx 24$, 
  even VLT spectroscopic capabilities are not sufficient. Photometric 
redshifts are the only technique to probe high-$z$ galaxies.  Moreover, has 
  it is demonstrated by several works done 
  in the past on photometric redshift (see for example 
 \citenum{bolzonella}), in order to 
  precisely sample the redshift domain between $z=0$ and $z=2$, 
    both optical and near infrared 
  photometric data are necessary, with S/N of 10 in each band. 
  Otherwise, part of the redshift range of the distorted sources 
  could be systematically  missed by the photo-$z$ machinery, which would
  bias the mass estimate of distant clusters. 
    In practice, only
  VLTs car go fast enough to probe these galaxies in all 
  $UBVRIJHK$ bands in a reasonable amount of time (see the 
results obtained with VLT/FORS+ISAAC on Fig.(\ref{zsource})). 
  In this respect it is 
  hard to HST and 4 meter class telescopes  to be competitive.  
   This is even more obvious if one wants to carry out a cluster weak lensing
  survey over a large cluster sample (see the EDICS cluster survey for 
  instance).  \\
Regarding dark clusters, the first priority is to test alternative 
   solutions, like very distant (almost 
  invisible) clusters of galaxies or projection effects. 
  VLTs are certainly very useful, in particular if dark clusters results 
  from projection effects along the line of sights. Deep optical and near infrared images of these fields can provide a detailed description of 
  source clustering which accumulate along each line of sight. This 
  contamination has been addressed recently by \citenum{whitelav02}) and 
it seems to be a serious concern.  Ultra deep multicolor observations 
  of each known dark cluster candidate can quickly test this possibility
  with VLTs.
\\
Finally, it is worth noticing that even for Poisson noise argument 
  the depth is also a strong requirement for very high-$z$ 
  clusters observations. Beyond
  $z \approx 1$, the galaxy number density of background sources 
   significantly decreases,  making the mass reconstruction more and more 
   noisy (and undersampled). Going to extremely deep imaging is the only way 
   to compensate the decreasing fraction of lensed population. 
  This critical point is also valid for photo-$z$, which will be more and
more demanding in observing time in order to keep 
  the signal-to-noise of photometric redshifts high enough.
   This is clearly beyond the capabilities of 
  4 meter class telescopes.
\\
There is indeed an alternative to depth in order to sample 
  accurately clusters and to investigate the  detailed 
  substructures in clusters of galaxies. It consists in using 
weak lensing in nearby clusters of galaxies (see \citenum{42sdss}). In that case, for a given 
   depth, one can probe small physical scales than for high-$z$ clusters. 
 In order to probe clusters on 
  megaparsec scales, this  alternative would favor wide field observations 
  rather than ultra deep VLT imaging.  However (\citenum{hoekstra01}) 
 pointed out that cosmic variance produced by all structure along the line of 
  sight may be a severe limitation of this approach for the nearest clusters.
\\
A general  limitation of VLTs is the rather small field of view 
  most faint imaging-spectrographs have. Even  
  the VLT/FORS intruments with of field of view of 7 arc-minutes
 cannot probe the periphery of $z \approx 0.5$ is one single shot.
 This then hampers 
  to assume that $\kappa=0$ at the border of the field. The mass-sheet 
degeneracy is still an issue for VLT and Keck telescopes, but likely not for 
  SUBARU/SUPRIME. 
\\ 
In conclusion, the VLTs are certainly the most precise telescope for a
  thorough investigations of clusters of galaxies, in 
  particular those at high redshift. The two weak points 
  are the image quality which cannot compete with HST and the small 
field of views they have.  On the first point, since galaxy angular size are
  significantly larger than the PSF and the pixel size, the capability of
  ground based telescopes to measure weak lensing down to a 
  limiting shear 
  of 1\%  is indeed excellent, even for VLT. Regarding
   the other point (which is even
 more dramatic for HST), clearly we should recommend to use SUPRIME at
SUBARU or VMOS at the VLT in the future. It would be extremely valuable to
  have similar wide field instrument for the near infrared. Even $J$
   and $H$
 bands, as it is proposed in the NIRMOS instrument design, would be sufficient.

\section{Cosmic shear}
Light propagation  over Gigaparsec distances and 
  across an inhomogeneous universe produces  
 weak lensing effects which accumulates along the line of sight of
 each ray bundle.
 Assuming structures formed from gravitational growth of
    Gaussian fluctuations, the shape and amplitude 
  of cosmological weak lensing as function of angular scale can be predicted
from Perturbation
  Theory.  To first order, the convergence $\kappa(\thetag)$ at
angular
position $\thetag$ is given by the line-of-sight integral
\begin{equation}
\kappa(\thetag)={3 \over 2} \Omega_0  \int_0^{z_s} n(z_s)  {\rm d}z_s
\int_0^{\chi(s)}  {D\left(z,z_s\right) D\left(z\right) \over
D\left(z_s\right)}  \delta\left(\chi,\thetag\right) \
\left[1+z\left(\chi\right)\right]  {\rm d}\chi
\end{equation}
where $\chi(z)$ is the radial distance out to redshift $z$, $D$ the
angular diameter distances, $n(z_s)$ is the redshift
distribution of the sources
  and   $\delta$ is
the mass density contrast responsible for the deflection at redshift
 $z$. Its amplitude at a given redshift  depends on the properties  of the
 (dark) matter power spectrum and its evolution with look-back-time.
\\
The cumulative weak lensing effects of
structures induce a shear field
  which is primarily related to the   power spectrum of the projected
mass density, $P_\kappa$.
   Its statistical properties can be analyzed using 
    several statistics on galaxy ellipticities, like the
shear top-hat variance (\citenum{me91,b91,k92}),
\begin{equation}
\langle\gamma^2\rangle={2\over \pi\theta_c^2} \int_0^\infty~{{\rm
d}k\over k} P_
\kappa(k)
[J_1(k\theta_c)]^2,
\label{theovariance}
\end{equation}
or others described in Mellier et al (\citenum{melspie1}, this conference)
 where $J_n$ is the Bessel function of the first kind.  $P_\kappa(k)$ is 
 directly related to the 3-dimension power spectrum of the dark matter 
   along the line of sight, $P_{3D}$.
\begin{equation}
P_\kappa(k)= {9 \over 4} \Omega_m^2 \ \int_0^\infty \ P_{3D}\left(
{k \over D_L\left(z\right)}, z\right) \ W(z,z_s) \ {\rm d}z \ ,
\label{p3d}
\end{equation}
where $ W(z,z_s)$ is an efficiency function which depends on the redshift 
  distribution of sources and lenses.
 Therefore,  in principle an inversion permit to reconstruct the
  3-dimension power spectrum of the dark matter from the 
  weak distortion field. \\
\subsection{Analysis and interpretation of the shear variance}
\begin{figure}[t]
\begin{center}
\includegraphics[width=10cm]{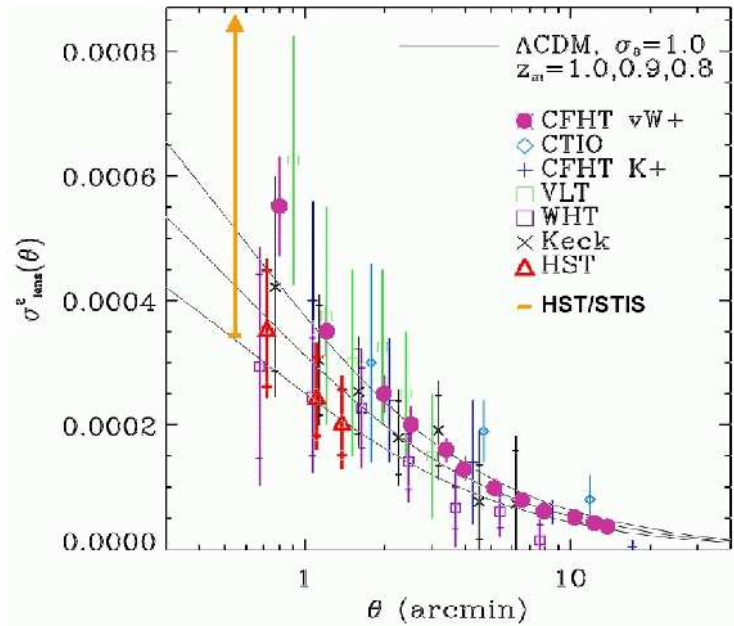}
\end{center}
\caption[]{Top hat variance of shear as function of angular scale from
  6 cosmic shear surveys (this plot is inspired by the 
  plot shown in (\citenum{bacon02})). The acronyms of the right refer to the 
  references reported in Table 2. 
}
\label{sheartop}
\end{figure}

\begin{figure}[t]
\begin{center}
\includegraphics[width=7.5cm]{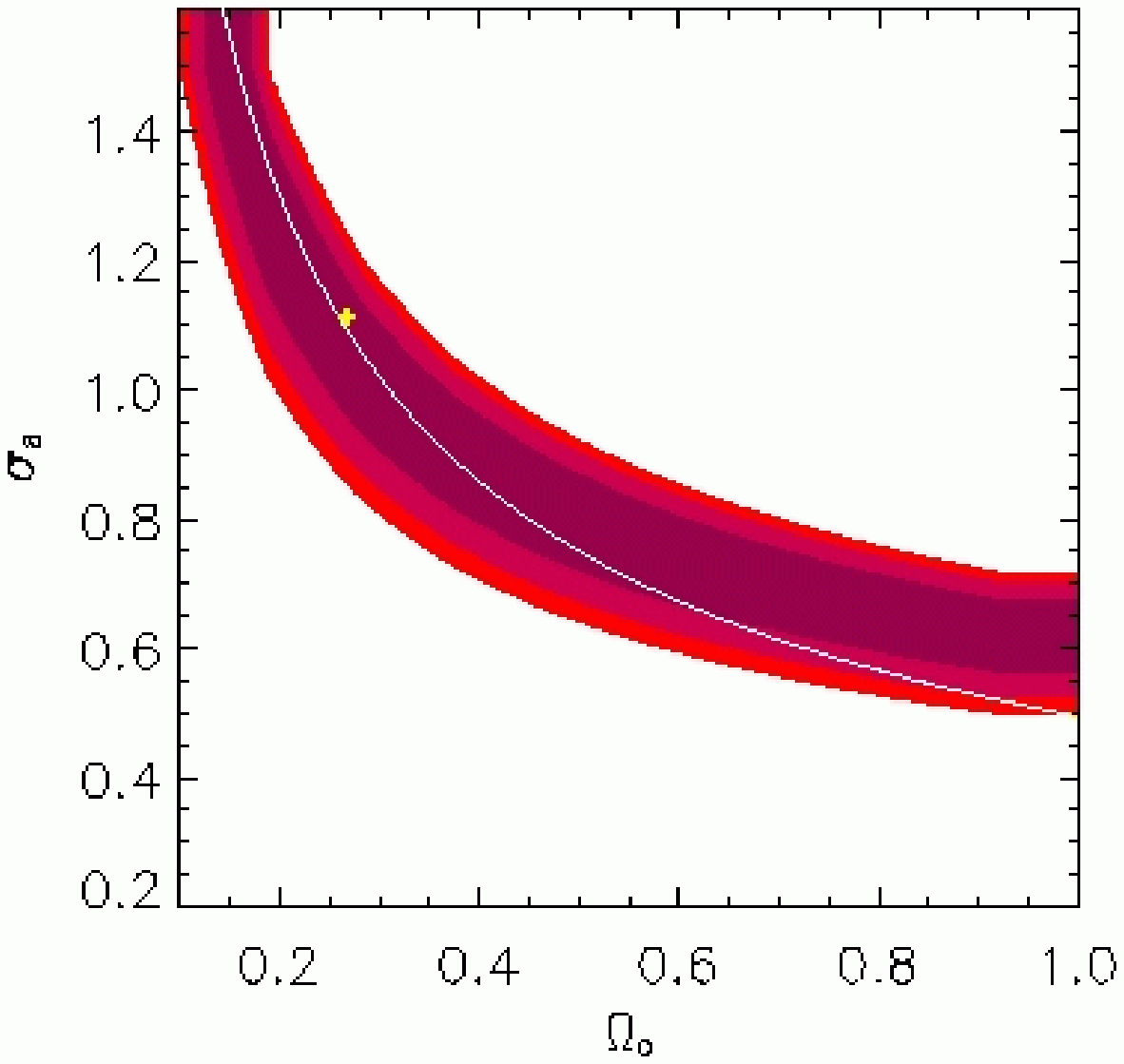}
\includegraphics[width=6.7cm]{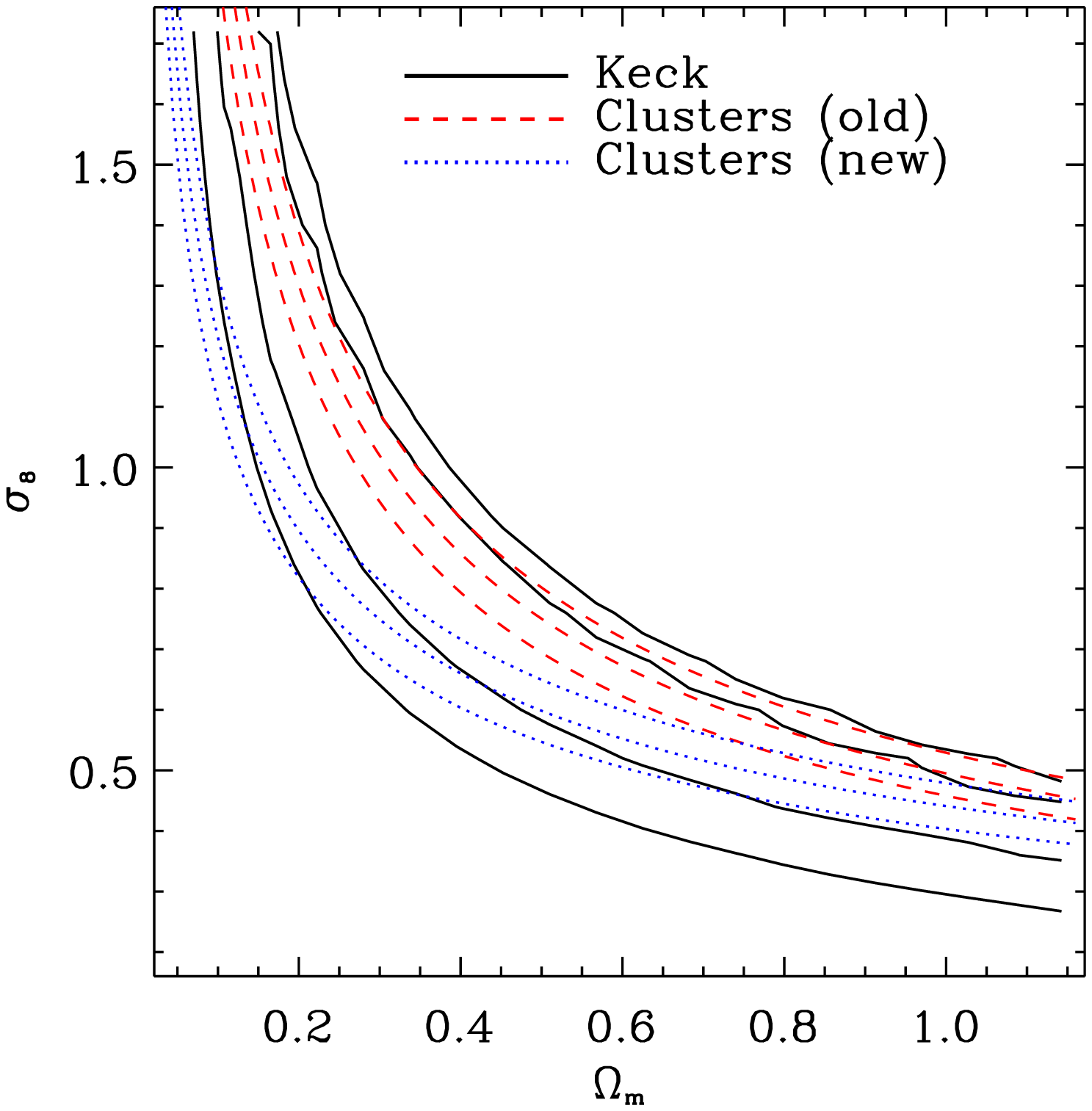}
\end{center}
\caption[]{$\Omega-\sigma_8$ constrains obtained with the 50 VLT 
 fields (left) by (\citenum{mao01}) and with the 170 Keck fields 
  (right) by (\citenum{bacon02}; courtesy D. Bacon).
  }
\label{cosmovlt}
\end{figure}

The amplitude
 of cosmic shear signal and its sensitivity to cosmology
  can be  illustrated in the fiducial case of a power
law mass power spectrum with no cosmological constant and
a background population at a single redshift
$z$. 
In that case, the variance of the convergence,  $<\kappa(\theta)^2>$,
  writes:
\begin{equation}
\label{eqvar}
<\kappa(\theta)^2>^{1/2} =<\gamma(\theta)^2>^{1/2} \approx 1\% \
 \sigma_8 \ \Omega_m^{0.75} \
z_s^{0.8
} \left({\theta \over  1'}\right)^{-\left({n+2 \over 2}\right)}  \ ,
 \
\end{equation}
where $z_s$ is the source redshift, $n$ is 
  the spectral index of the power spectrum of density fluctuation and
   $\sigma_8$ the power spectrum normalization.  
  Several teams  
   succeeded to get a significant signal
 (see Table \ref{tabcs}).
  Since each group used different telescopes,
  adopted different observing strategies and
 used different data analysis techniques,  one can figure out
  their reliability by comparing these surveys.
\begin{table}
\begin{center}
{\small
\caption{Present status of cosmic shear surveys with published
results.
}
\label{tabcs}
\begin{tabular}{lcccl}\hline
Telescope& Pointings & Total Area & Lim. Mag. & Ref.. \\
\hline
CFHT & 5 $\times$ 30' $\times$30'& 1.7 deg$^2$ & I=24. &
\citenum{vwal00}[vWME+]\\
CTIO & 3 $\times$ 40' $\times$40'& 1.5 deg$^2$ & R=26. &
\citenum{wit00a}[WTK+]\\
WHT & 14 $\times$ 8' $\times$15'& 0.5 deg$^2$ & R=24. &
\citenum{bacon00}[BRE]\\
CFHT & 6 $\times$ 30' $\times$30'& 1.0 deg$^2$ & I=24. &
\citenum{kais00}[KWL]\\
VLT/UT1 & 50 $\times$ 7' $\times$7'& 0.6 deg$^2$ & I=24. &
\citenum{mao01}[MvWM+]\\
HST/WFPC2 & 1 $\times$ 4' $\times$42'& 0.05 deg$^2$ & I=27. &
\citenum{rhodes01}\\
CFHT & 4 $\times$ 120' $\times$120'& 6.5 deg$^2$ & I=24.
&\citenum{vwal01}[vWMR+]\\
HST/STIS & 121 $\times$ 1' $\times$1'& 0.05 deg$^2$ & V$\approx 26$
& \citenum{hammerle01}\\
CFHT & 5 $\times$ 126' $\times$140'& 24. deg$^2$ & R=23.5& \citenum{hoek01a}
\\
CFHT & 10 $\times$ 126' $\times$140'& 53. deg$^2$ & R=23.5& \citenum{hoekstra02}
\\
CFHT & 4 $\times$ 120' $\times$120'& 8.5 deg$^2$ & I=24.&
\citenum{pen01}\\
HST/WFPC2 & 271 $\times$ 2.1 $\times$ 2.1 & 0.36 deg$^2$ & I=23.5 & \citenum{ref02}
 \\
Keck+WHT & 173 $\times$ 2' $\times$ 8' & 1.6 deg$^2$ & R=25 & \citenum{bacon02} \\
 & +13 $\times$ 16' $\times$ 8' &  &  &  \\
 & 7 $\times$ 16' $\times$ 16' & &  &  \\
\hline
\end{tabular}
}
\end{center}
\end{table}
Figure \ref{sheartop} show that they are all
   in very good
agreement\footnote{ The Hoekstra et al (\citenum{hoek01a}) data are missing
   because depth is
  different so the
sources are at lower redshift and the amplitude of the shear
  is not directly comparable to other data plotted}.  This  is
  a convincing demonstration that the
   correlation of ellipticities do not only results from 
  unexpected systematic effects.
The cosmological origin of the coherent distortion signal detected 
   by all these surveys has been confirmed by  
  decomposing the distortion field  into  E- and B- modes.
   The E-mode is a gradient term which 
 contains signal produced by gravity-induced distortion. 
 The B-mode is a pure curl-component, so it
  only contains intrinsic ellipticity correlation or
systematics residuals. Both modes have been extracted using the
aperture mass statistics by van Waerbeke et al (\citenum{vwal01},
 \citenum{vw02}) 
 \ and
 Pen et al (\citenum{pen01}), in the VIRMOS-DESCART\footnote{
For the VIRMOS spectroscopic survey, see (\citenum{olf02}). 
  For the VIRMOS-DESCART survey see http://terapix.iap.fr/Descart/}\footnote{Data of the VIRMOS-DESCART survey were processed and analyzed at the
 TERAPIX data center: http://terapix.iap.fr}
     survey, as well as
  by Hoekstra et al (\citenum{hoekstra02}) in the Red Cluster Sequence survey.
 In both samples, the E-mode dominates the signal,
  which strongly supports the
   gravitational origin of the distortion. 
\\
The comparison of the top-hat shear variance 
   with some realistic cosmological models has been done for these 
surveys. Although some cosmological models are ruled by these 
  data, as  expected from Eq.(\ref{eqvar}) 
   the degeneracy between $\Omega_m$ and
   $\sigma_8$   hampers
  a strong discrimination among most popular cosmological models.
    The present-day constraints resulting from these studies, 
  including the VLT and Keck samples (see Figure(\ref{cosmovlt})), 
   can be summarized as follows
   (90\% confidence level):
\begin{equation}
0.05 \le \Omega_m \le 0.8 \ \ \ \ {\rm and} \ \ \ \ 0.5 \le \sigma_8
\le 1.2 \ ,
\end{equation}
 and, in the case of a flat-universe with $\Omega_m=0.3$,  
   we infer $\sigma_8 = 0.9 \pm 0.1$ (68\% c.l.).
\\
\subsection{Analysis of non-Gaussian features in cosmic shear signal}
\begin{figure}[t]
\begin{center}
\includegraphics[width=8.7cm]{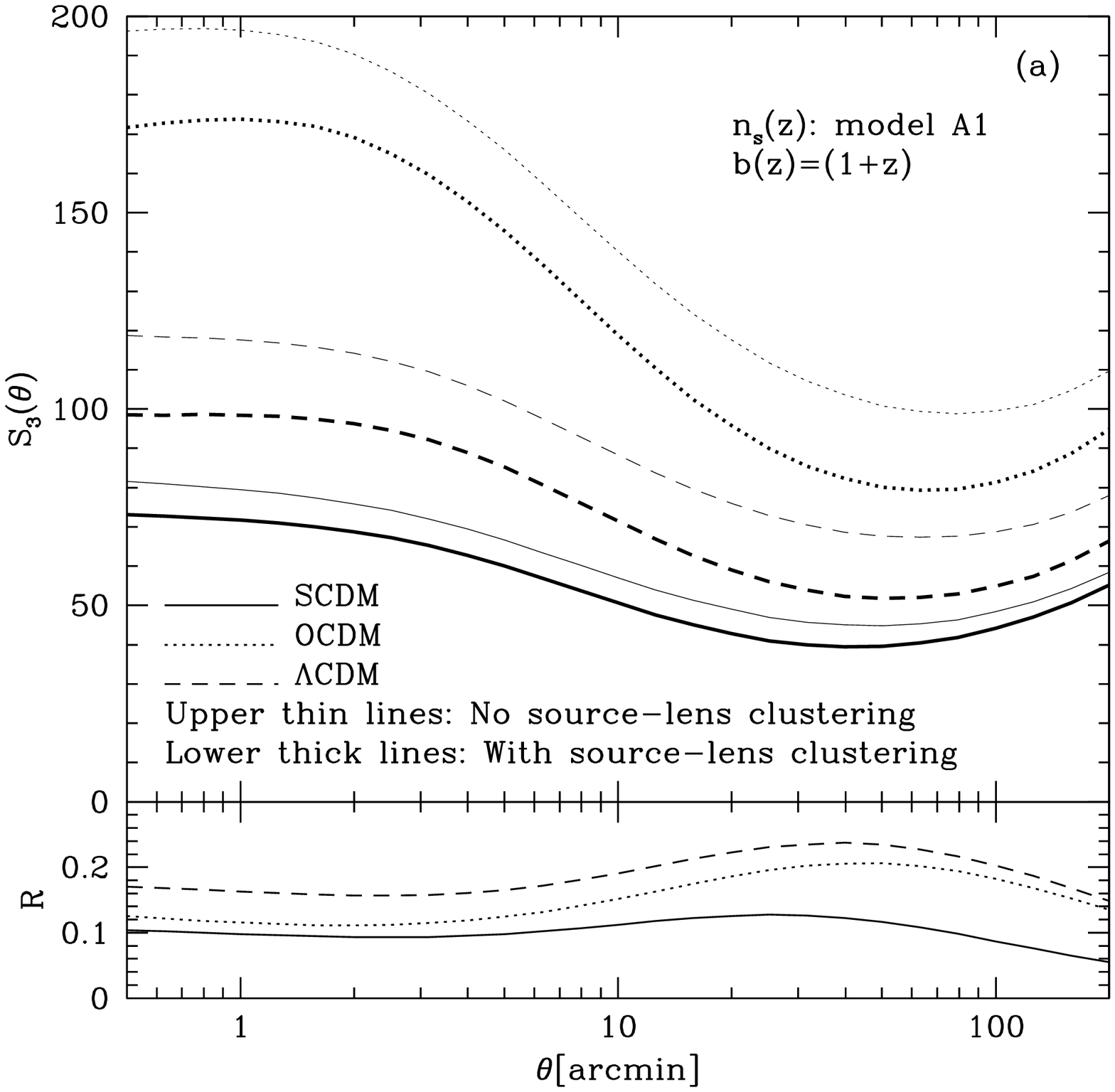}
\includegraphics[width=6.0cm]{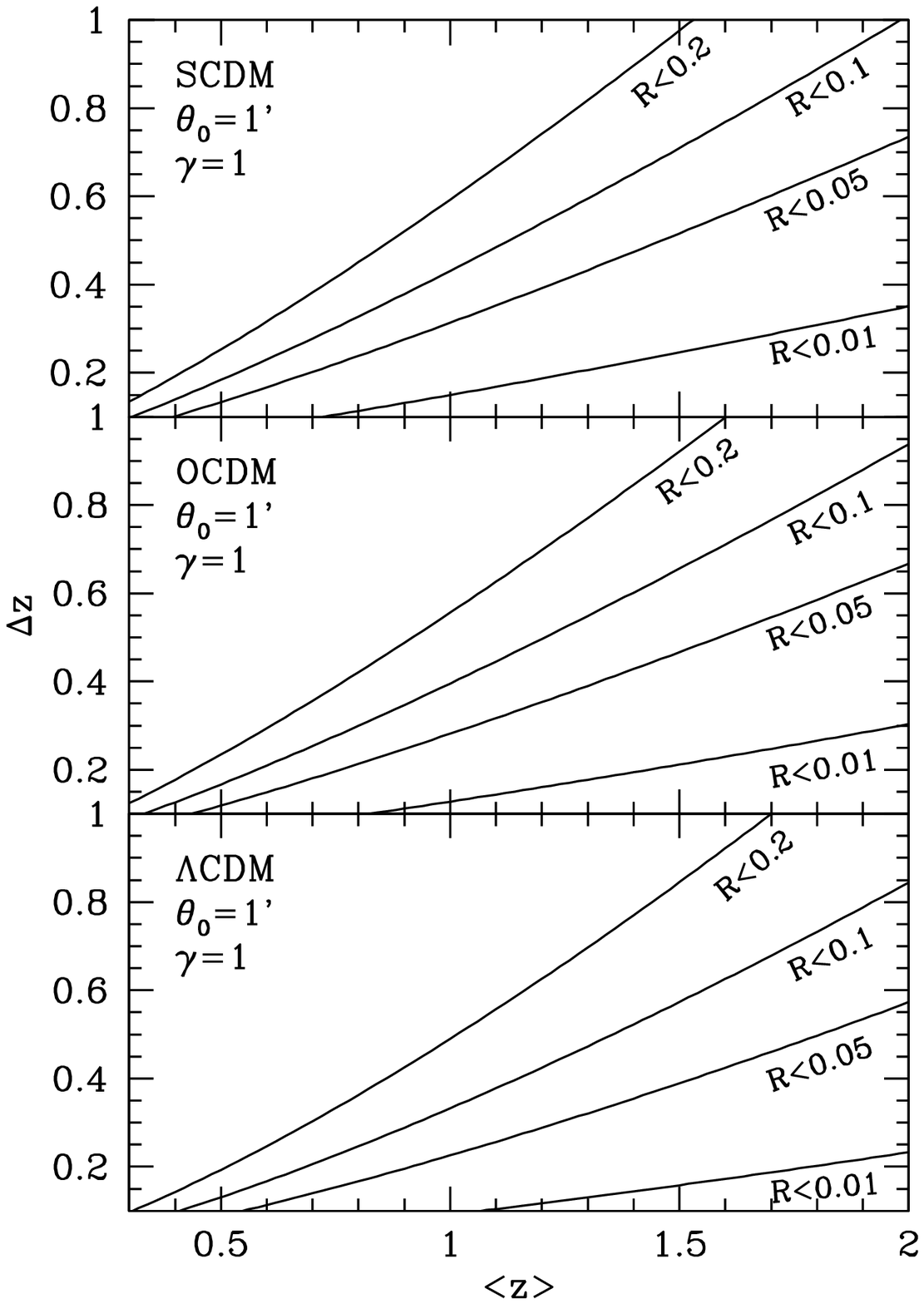}
\end{center}
\caption[]{ Sensitivity of the convergence skewness to source-lens clustering.
  The top-left panel shows how the amplitude of the skewness varies with 
  source-lens clustering for three cosmological models. The lower plot 
  gives the relative variation. In all cases, the skewness decreases by
  a factor larger than 10\%,  a serious limitation to precision 
cosmology with cosmic shear.  To solve this issue, one needs
  to get the redshift of source-lensesr. On the right panel, the width of 
  the source distribution is plotted as function of the 
 averaged source-redshift.
  In order to minimize the clustering effect,
  the source distances must be large as possible and 
spread over a narrow redshift range. Clearly, the accuracy of 
  photometric redshift must be better than 10\%. This goal can be
  achieved if $UBVRIJ$ and $K$ band data are obtained for most lensed 
  galaxies, with a photometric accuracy of about 10\% in each filter.
  (From \citenum{hamana00})}
\label{hamana}
\end{figure}

 Higher order statistics, like  the skewness
of the convergence, $s_3(\theta)$, can also be computed.
  These statistics are sensitive to non Gaussian
features in the projected mass density field produced by non-linear 
   systems like
    massive clusters or compact groups of galaxies (\citenum{bern97},
 \citenum{jain97}).  
As for the simple fiducial models discussed in the previous section, 
  one can express also  $s_3(\theta)$:
\begin{equation}
\label{eqs3}
s_3(\theta)={\langle\kappa^3\rangle\over \langle\kappa^2\rangle^2}
\approx 40 \
  \Omega_m^{-0.8} \ z_s^{-1.35}  \ .
\end{equation}
 Therefore, in principle the
      degeneracy between $\Omega_m$ and $\sigma_8$ can be  broken
   when both the variance and the skewness of the convergence
  are measured.
\\
Unfortunately,  the measurements of $s_3(\theta)$ 
   is presently hampered by a number a practical difficulties which are not yet
    fixed.  In particular, the masking process  
   which masks bright stars and defects on CCD images degrades the quality 
  of  the mass reconstruction. The 
  reliability of its skewness as well as its cosmological interpretation   
  are therefore seriously questioned.
 Bernardeau, van Waerbeke \& Mellier (\citenum{bern02a})  have
proposed an alternative method using some specific patterns in the shear
three-point function.
  Their  detection strategy based on their method has
been tested on ray tracing simulations and
  turns out  to be robust, usable in patchy catalogs, and quite
insensitive to the topology of the survey.
 They have recently used the analysis
  of the  3-point correlations function on the VIRMOS-DESCART
data.  Their results  show a
2.4$\sigma$ signal over four independent angular bins
 corresponding to angular scales between  $2$ and $4$ arc-minutes.
The amplitude and the shape of the signal are consistent
with theoretical expectations obtained from ray-tracing simulations.
This result supports
the idea that the measure corresponds to a  cosmological signal due to
the gravitational instability dynamics. Although the
   errors
 are  still large to permit secure conclusions, one clearly sees that
   the amplitude and the shape of the 3-point correlations function
   match the most likely cosmological models.  Remarkably, the
  $\Lambda$CDM scenario perfectly fit the data points.
\\
The Bernardeau et al (\citenum{bern02b}) result is the
first detection of non-Gaussian features in a cosmic shear survey and
  it opens the route to break the $\Omega_{{\rm m}}-\sigma_8$ degeneracy.
 However, there are still some caveats which may be considered
seriously.
 One difficulty is the source-lens  clustering which could
   significantly perturb high-order statistics
 (Hamana et al 2002 (\citenum{hamana00})). Fig.(\ref{hamana}) shows that 
  source clustering may change the amplitude of the skewness of the
  convergence by 20\%.
 If so, multi-lens plane cosmic shear analysis will be necessary
  which implies a good knowledge of the redshift distribution.
  According to Hamama et al (Fig. \ref{hamana}), the source clustering 
  contamination decreases with source redshift, so that  
  this issue is less critical when lensed galaxies are
a very high-$z$.   For sources at redshift one, the effect of source clustering 
  can drop down to less than 5\% if the  sources are within a narrow 
  range in redshift. It implies that source redshift  must be measured
with an accuracy better than $0.2$ at redshift one, and better than 
  0.05 for source at redshift 0.5. When looking at the Bolzonnela et al
   plot on photo-$z$ (see \citenum{bolzonella}), this accuracy demands
   both optical and near infrared photometric data, with a 
  photometric precision of about 10\% in all filters.
\subsection{Issues and prospects for VLTs}
Although there are still a lot of  technical issues
  that must be addressed prior to go toward high precision cosmology 
  with cosmic shear, there are some immediate points which turn out to
  be important for the near future and where VLTs can help. 
  As for clusters, let us present them 
   by the following questions:
\begin{itemize}
\item What is the redshift distribution of sources?
\item Can be have an idea of source clustering for skewness?
\item How can we minimize cosmic variance?
\item How high order statistics are contaminated by $E$ and $B$ modes?
\item Are VLT competitive with respect to wide field 4-meter telescopes for 
cosmic shear?
\end{itemize}
These issues express the sensitivity of the variance and the skewness
  to source redshift, as shown in Eq.(\ref{eqvar}, \ref{eqs3})
  and to the source-lens clustering, as pointed out by (\citenum{hamana00}).  
On the other hand, the strength of VLT with respect to 
  other telescopes can also be discussed by looking at the 
  signal-to-noise ratio of cosmic shear surveys.
  Adopting the 
  null hypothesis that no gravitational weak lensing signal is present, 
  we can predict the expected limiting shear  
  if only shot noise and sampling are taken into account (in particular,
   we assume there is no systematic residual errors associated to 
  image processing or PSF corrections as those discussed later). 
  For a $3\sigma$ shear variance limiting detection it writes :
\begin{equation}
\label{survey}
<\gamma(\theta)^2>_{limit}^{1/2} = 1.2\%  \ \left[{A_T \over 1
\ deg^2}\right]^{-{1 \over 4}} \times \left[{\sigma_{\epsilon_{gal}}
\over 0.4} \right] \times
 \left[{n \over 20 \ gal/arcmin^2}\right]^{-{1 \over 2}} \times
 \left[{\theta \over 10'}\right]^{{-{1 \over 2}}} \ ,
\end{equation}
where $A_T$ is the total area covered by the survey, $\sigma_{\epsilon_{gal}}$ 
 is the intrinsic ellipticity dispersion of galaxy and $n$ the galaxy number 
  density of the survey. 
\\
The sensitivity to $A_T$ favors CCD panoramic cameras
   on  4 meter telescopes against VLTs with small field of view.
  However, the sensitivity to $n$ compensates this weak point. Furthermore, 
  since VLT are faster photon collectors than other ground based telescopes, 
  they can observe much more uncorrelated fields, which can be 
spread over a very large sky area, which then lower the cosmic variance 
  significantly. Therefore, 
  exactly as for clusters of galaxies, the need for VLT on cosmic shear 
 projects is primarily the need for very accurate photometric redshifts
  and for very deep exposure to increase the galaxy number density. 
  All conclusions relevant for clusters are therefore valid for cosmic shear.
 In addition,  it is important to  benefit from the speed of VLT
  to define an observing strategy based on a sparse sampling 
   using a large number of fields rather
  than a very large total and compact field of view.  
\section{Summary and conclusions}
\begin{figure}[t]
\begin{center}
\includegraphics[width=10cm]{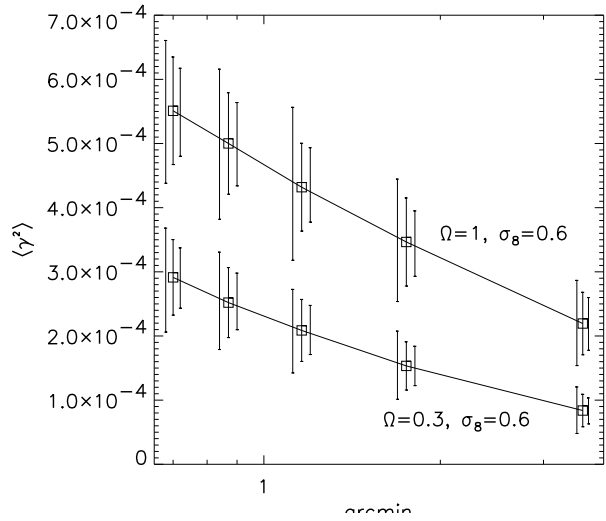}
\end{center}
\caption[]{Illustration of the signal-to-noise of the variance
of the shear for 100, 150 and 300 VLT/FORS1 fields. One can see
  that for 300 hundred fields, the two models will be
disentangled with a 3$\sigma$ confidence level.
  }
\label{nextsurvey}
\end{figure}

The very large aperture of VLTS make these telescopes the most competitive 
 for all scientific goals which demand very deep or very fast 
  capabilities.   By increasing limiting magnitudes, VLTs permit to
  increase the galaxy number density of lensed galaxies, which in return 
 drops the Poisson noise,  improves the spatial resolution of mass maps 
  and increases the shear signal by increasing the redshift of sources. 
  As we discussed in previous sections, the knowledge of redshift of sources 
  as well as their clustering will be more and more critical in the 
future for high precision cosmology.  Both optical and 
  near infrared capabilities  will therefore be necessary.  This requirement 
  is hard to reconcile with high depth unless unrealistic 
  amount of telescope time 
   is spent on weak lensing and cosmic shear studies.  Indeed VLTs are
  in this respect the best instruments, and a join FORS1+ISAAC-like 
   configuration looks  promising for future surveys, provided 
   the near infrared field of view is large enough.  
\\
The speed capabilities of VLTs is almost of equal importance. As 
 (\citenum{mao01})
   and (\citenum{bacon02}) showed, 10 meter class telescopes can spend only 
  20 mn per field instead of one hour for 4 meter class telescopes. This
  advantage can be used to sample in a clever way the sky and make 
  a sparse coverage of a huge field.  Maoli et al (\citenum{mao01}) 
  spread 50 6'$\times$6' FORS1 
field overs 2000 deg$^2$ and Bacon et al
 (\citenum{bacon02})  spread 170 fields in the same 
  way.  Clearly, one could envision to observe at least 1000 fields
  with VLT.  Figure (\ref{nextsurvey}) show the gain when a survey 
  is based on  
  100, 150 or 300 fields...
  One can  see the potential to constrain cosmological models, with 
  in addition 
   accurate photometric redshift and therefore 
   clustering informations for each field. This strategy can be also used 
  to probe in a clever way the power spectrum on the projected mass density.
  However, there is not yet a clear quantitative evaluation of the 
  optimum efficiency of such a survey as function of its topology 
  and the sampling strategy. This point has to be clarified.
\\
The  weakness of VLTs  is definitely the rather  small field of view.
  SUBARU/SUPRIME with of field of view of 24' turns out to 
  be the most competitive configuration 
    (Miyazaki 2002, (\citenum{subashear})). 
  For weak lensing studies of distant clusters of galaxies, the field 
  of view of  is not a serious issue since 1 Megaparsec roughly corresponds
  to 5' .    Therefore,  VLT are definitely superior for weak lensing 
  analysis of distant clusters of galaxies.
    In contrast, as Eq.(\ref{survey}) 
   shows, this may be a limitation for cosmic shear surveys.  The 
  incoming VMOS camera on the VLT, which will have a 15' field of view 
  will be a very competitive instrument for cosmic shear, provided it 
is demonstrated that is image quality is not degraded at the boundary of the field.  The need for a wide field near infrared instrument with similar
field of view as optical camera is clear. The $K$-band is not really
necessary since even $J$ and $H$ bands are sufficient to get precise 
  photometric redshifts. With such an optical+NIR configuration, 
  a cosmic shear survey covering at least 300 hundred uncorrelated fields
   with photo-$z$ for all galaxies will produce the most 
   precise data set for high precision cosmology using gravitational 
lensing tools.
\\

\section*{Aknowledgements}
 We thank D. Bacon, M. Bartelmann, D. Clowe, R. Ellis, B. Fort, H. Hoekstra, 
  L. King,  M. Lombardi,  A. Renzini and A. R\'efr\'egier 
 for useful discussions on the potential of very large telescopes for
  weak lensing. We thank S. White and the EDICS team and in particular 
  D. Clowe for providing the mass map of cl1232-1250 prior to
  publication and D. Bacon for providing 
  authorization to publish a figure of his paper.  This work was supported by the TMR Network
``Gravitational
 Lensing: New Constraints on
Cosmology and the Distribution of Dark Matter'' of the EC under contract
No. ERBFMRX-CT97-0172. 


\end{document}